# Depth-controlled Bessel beams

Angelina Müller, Matthias C. Wapler and Ulrike Wallrabe
University of Freiburg – IMTEK, Germany
E-mail: wallrabe@imtek.uni-freiburg.de

*Abstract*— We present a ring aperture with independently switchable segments for the three-dimensional control of quasi propagation invariant beams. We demonstrate that our liquid crystal design concept preserves coherence and generates the Bessel beam structure.

## I. Introduction

Bessel beams are quasi propagation invariant beams coming from a self-interfering conical wavefront, e.g. created by an axicon. Featuring an extended focal zone and the property of self-reconstruction, they are used in many different fields, e.g. for optical imaging or material processing [1,2]. To control the lateral region of this – theoretically infinite – focal zone, we develop a ring aperture based on a nematic liquid crystal cell. Liquid crystal (LC) devices are widely used in different applications, e.g. in spatial light modulators which can also generate Bessel beams [3]. While the long-term aim is an integrated miniaturized system, here, we present the proof of concept using a LC cell with three segments and a commercial axicon, as schematically shown in Fig. 1.

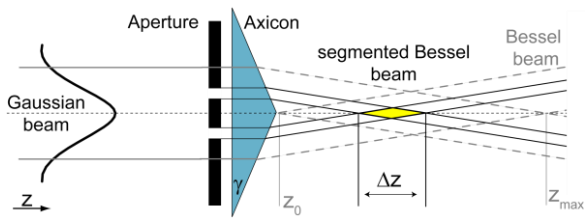

Fig. 1. Working principle of the system. The highlighted area Δz indicates the segmented Bessel beam. The dotted lines indicate the Bessel beam without an aperture.

## II. Fabrication

For the fabrication, we used standard MEMS thin-film and lamination technology as shown in Fig. 2. We start with an ITO-coated, 0.7 mm thick Pyrex substrate with a sheet resistance of 20 Ω/sq. First, we deposited the contact pads (100 nm Ti / 100 nm Pt) using PVD and lift-off and then structured the ITO electrodes using photolithography and wet etching. The LC cavity was created using a 60 µm thick Ordyl layer by lamination and photolithography. After dispensing and rubbing the LC alignment layer, we flip-chip bonded both sides and inserted the LC (CB5). We designed the cavity in such a way that the droplet will flow inside via capillary forces. As we cannot guarantee that the fiber-coupled laser is still polarized, we finally added linear polarizers with 90° offset to each other (see Fig. 3) and mounted the device in a PCB.

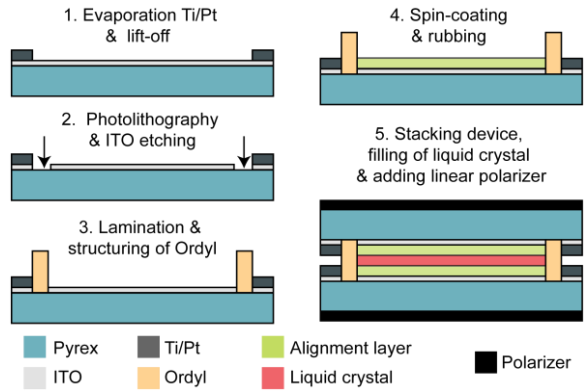

Fig. 2. Fabrication steps of the LC ring aperture device.

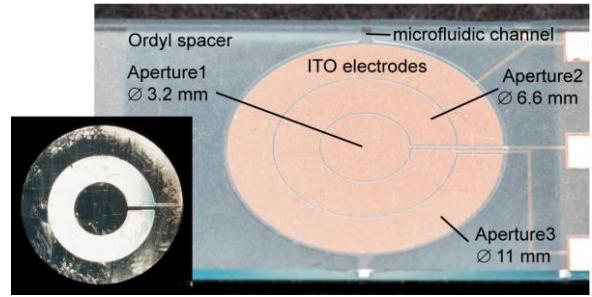

Fig. 3. Photograph of the unfilled LC ring aperture with three individually addressable channels. Inset: Assembled ring aperture with actuated aperture 2.

## III. Characterization

We performed our measurements using a HeNe laser with $\lambda = 632.8$ nm, $1/e^2$ radius r = 4 mm and a UV fused silica axicon with an angle γ = 1° (see Fig. 1). The resulting first minimum of the Bessel beam ($r_{core}$) is 30.36 µm and the maximum length of the focal zone $z_{max}$ is 501.5 mm. We measured the beam profile using a camera with 1.67 µm pixel pitch, which is mounted on a linear translation stage in the propagation direction. Furthermore, we used a signal of 500 Hz and 4 $V_{rms}$ to drive the LC device.

To compare the measurements with the theoretical expectations, we simulated the beam propagation through the ring aperture using fast Fourier transformations (FFT) similar to [4] and used the measured axicon profile together with a Gaussian approximation for the incoming laser beam profile.

In Fig. 4 we show the simulation and measurement results of the whole Bessel beam without the ring aperture and the segmented Bessel beams created with the ring aperture. Up to some variations, the measurements agree well with the simulations. The cross-sectional images in Fig. 4 (bottom, right) clearly show the characteristic ring pattern of the Bessel beam, however, with some unevenness due to wavefront errors at positions (c) and (d), i.e. at large ring radii.

To investigate the results more quantitatively, we show the normalized core intensity of the Bessel beam along its propagation direction in Fig. 5. The segmented beams follow the Bessel beam without an aperture very closely in the corresponding region. They start slightly later than the shaded purely geometrical regions and show some oscillations coming from aperture effects. The measured beam profile varies slightly from the simulation, as we assumed a perfect Gaussian profile, which the real illumination source does not fully provide. Still, the segmentation is clearly visible. However, the oscillations are highly suppressed, probably due to the smoother edges of the fabricated apertures.

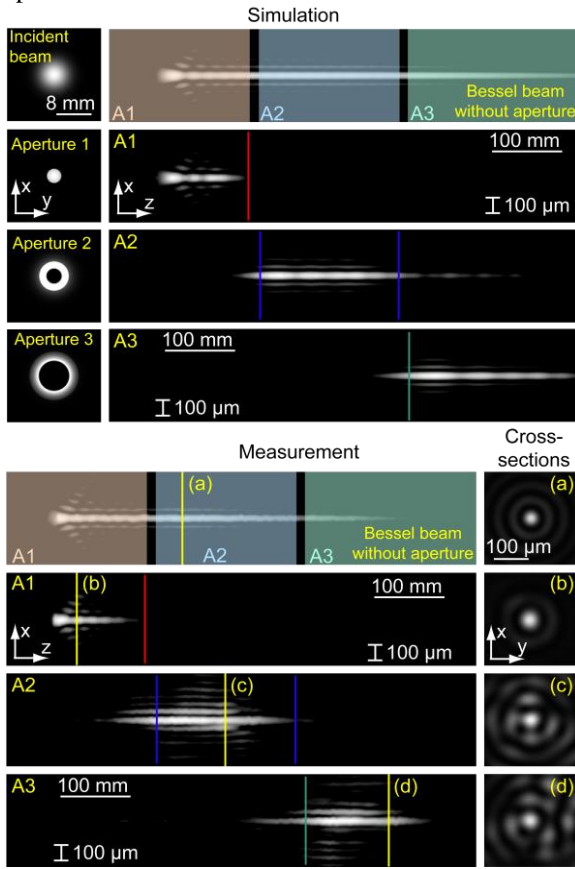

Fig. 4. Simulated (top) and measured (bottom) intensity profile of the Bessel beams (logarithmically scaled) with and without the ring apertures. The shaded areas indicate the purely geometrical Bessel beam regions.

## IV. CONCLUSIONS

We have demonstrated a new device to control the depth of Bessel beams using liquid crystal technology. Already with a large device, the wavefront error is small, and the measurements show good agreement with the simulations. In the future, we plan to further miniaturize the system and integrate it with micro axicon arrays, which will further reduce the wavefront error.

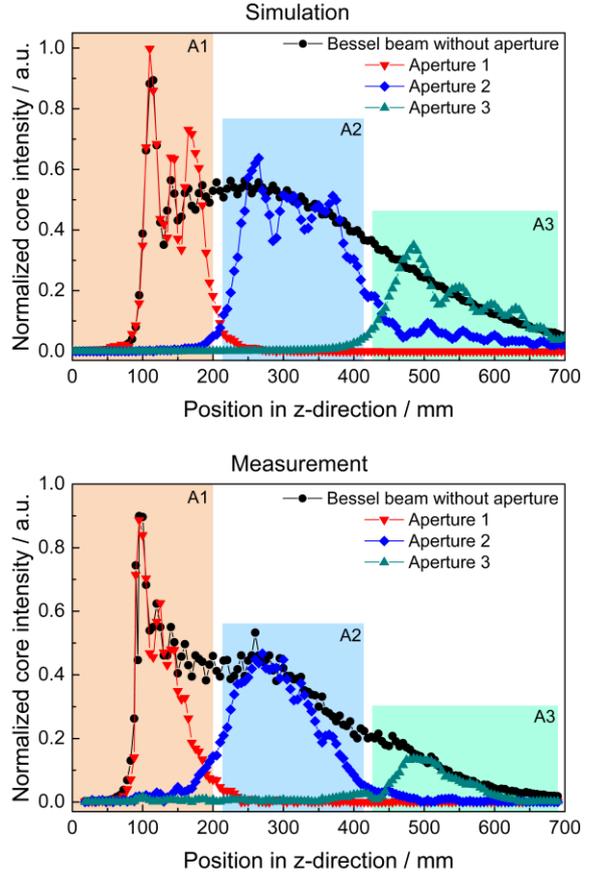

Fig. 5. Core intensity along the propagation direction of the simulated (top) and measured (bottom) beam with and without ring apertures, normalized to the incident beam. The shaded areas indicate the regions obtained from purely geometrical considerations.

ACKNOWLEDGMENTS

This work was supported by the BrainLinks-BrainTools Cluster of Excellence funded by the German Research Foundation (DFG, grant no. EXC 1086).